\documentclass[aps,pre,twocolumn,superscriptaddress]{revtex4-1}

\usepackage{graphicx}
\usepackage{amsmath}
\usepackage{enumerate}

\begin{document}


\title{Asymmetric Rotational Stroke in Mouse Node Cilia during Left--Right Determination }


\author{Atsuko Takamatsu}
\affiliation{Department of Electrical Engineering and Bioscience, Waseda University, Shinjuku-ku, Tokyo, 169-8555, Japan}

\author{Takuji Ishikawa}
\affiliation{Department of Bioengineering and Robotics, Tohoku University, Sendai, Miyagi, 980-8579, Japan}

\author{Kyosuke Shinohara}
\affiliation{Developmental Genetics Group, Graduate School of Frontier Biosciences, Osaka University, Suita Osaka 565-0871, Japan}

\author{Hiroshi Hamada}
\affiliation{Developmental Genetics Group, Graduate School of Frontier Biosciences, Osaka University, Suita Osaka 565-0871, Japan}


\date{\today}

\begin{abstract}
Clockwise rotational movement of isolated single cilia in mice embryo was investigated in vivo. The movement generates leftward fluid flow in the node cavity and plays an important role in left--right determination. The leftward unidirectional flow results from tilting of the rotational axis of the cilium to the posterior side. Because of the no-slip boundary condition at the cell surface, the upper stroke away from the boundary generates leftward flow, and the lower stroke close to the boundary generates slower rightward flow. By combining computational fluid dynamics with experimental observations, we demonstrate that the leftward stroke can be more effective than expected for cases in which cilia tilting alone is considered with the no-slip condition under constant driving force. Our results suggest that the driving force is asymmetric and that it is determined by the tilting angle and open angle of the rotating cilia. Specifically, it is maximized in the leftward stroke when the cilia move far from the boundary.
\end{abstract}


\maketitle

Cilia or flagella are widely found in biological systems from bacteria to mammals \cite{Bray00}. Their movements relate directly or indirectly to their own biological functions, e.g., protozoa swimming, coordinated beating of cilia in epithelium, and rotational movement of mono-cilia in mouse nodes. This study specifically examined the third, mouse node cilia, which are involved in left--right (LR) determination \cite{Marshall06,Shiratori06}.

\begin{figure}[h]
	\includegraphics[width=0.5\textwidth]{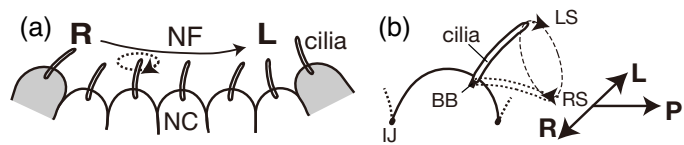}%
	\caption{Node cilia. L, R, and P respectively denote left, right, and posterior directions in a node: NC, node cell; 
NF, node flow; LS, leftward stroke; RS, rightward stroke;
BB, basal body; IJ, intercellular junction.
\label{fig:nodalcilia}}
\end{figure}

A small cavity, called a node, is formed in a mouse embryo at an early developmental stage (7.75 days after fertilization). The node cavity, which is filled with extra embryonic fluid, is constructed with a few hundred cells, each of which protrudes a single motile cilium (Fig. \ref{fig:nodalcilia}(a)). The cilium consists of nine doublet microtubules without a central pair, called 9+0 or a primary cilium, differently from the other ordinary motile cilia consisting of nine doublets with central pairs of microtubules (9+2). Cilia generate a rotational or beating movement (mainly moving back and forth), respectively, in 9+0 or 9+2 cilia, by sliding motion of dynein motor proteins along the microtubules. It is particularly interesting that the rotational movement is almost always clockwise when viewed from the ventral side, at least in mouse, rabbit, and medaka \cite{Okada05}. 
In this paper, we show that the driving force of the rotational movement can be rotationally asymmetric in contrast to the fact the 9+0 has structural rotational symmetry.

The clockwise rotational movement by the 9+0 cilia serves an important role in LR symmetry breaking. The movement generates leftward laminar flow of up to 1--10 $\mu m$/s because the rotational axis of the cilium tilts to posterior side. Because of a no-slip boundary (NSB) condition on the cell surface, upper leftward strokes (LS), which are less affected by NSB, generate faster leftward flow, and a lower rightward stroke (RS), which is directly affected by the boundary, generates a slower rightward flow (Fig. \ref{fig:nodalcilia}(b)). Results reported herein show that LS is more effective than expected before, when only the NSB condition was considered under a constant driving force of the rotational movement \cite{Marshall06}. The result was derived by analyzing rotational movement of isolated single cilia using mutant mice, then combined with computational fluid dynamics.

\begin{figure}
	\includegraphics[width=0.45\textwidth]{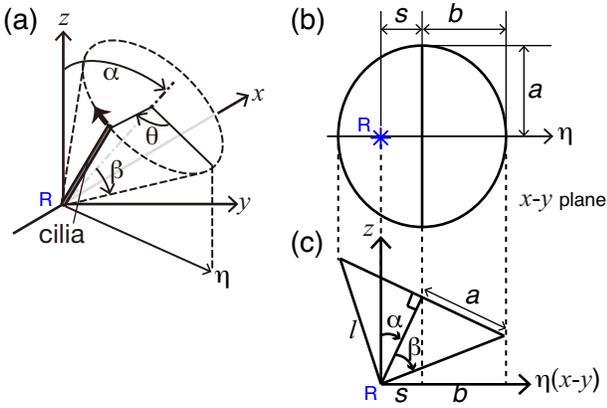}%
	\caption{(color online) (a) Setup of cilia, (b) top view (x--y plane projection), (c) side view (z-$\eta$ projection), {\bf R} denotes the cilia root position.
	\label{fig:ciliaparameter}}
\end{figure}

\begin{figure}
	\includegraphics[width=0.5\textwidth]{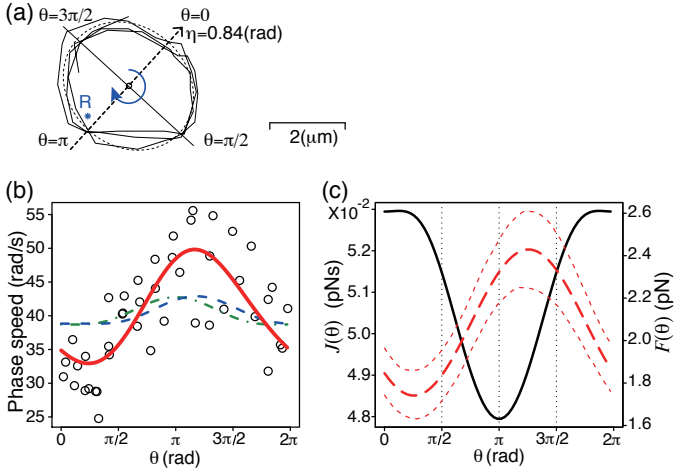}%
	\caption{(color online) Example of estimation of $F(\theta)$. (a) Projected trajectory, where cilia parameters were estimated as $\alpha=0.37$, $\beta=0.61$, $l=3.2$ ($\mu$m). (b) Phase speed fitted by 
Eq. (\ref{eq:rotationEq}) using
the three models. Green dotted and dashed, blue dashed, and red lines respectively represent fitting results by model (1--3). Data numbers of phase-speed $n=44$. (c) $F(\theta)$ calculated using model (3). Black solid lines denote $J(\theta)$, red dashed lines denote $F(\theta)$.
Red short dashed lines represent a 95\%- confidence interval associated with $F(\theta)$, which is estimated from the dataset of phase and phase speed (plots of Fig. \ref{fig:simulation}(b)) using the bootstrap method \cite{Efron86}. 
\label{fig:simulation}}
\end{figure}

\begin{figure}
	\includegraphics[width=0.52\textwidth]{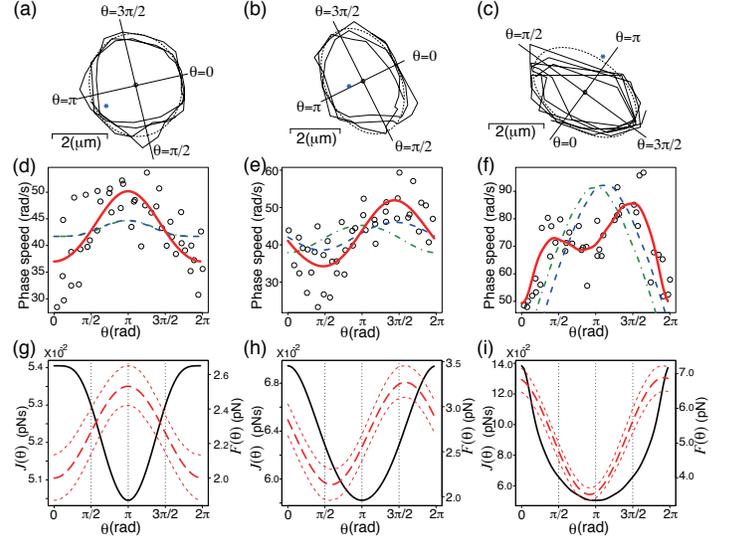}%
	\caption{(color online) Calculation examples of $F(\theta)$. (a--c) Projected trajectory. Cilia parameters were estimated respectively as $\alpha$=0.30, 0.10, 0.68, $\beta$=0.58, 1.24, 0.86, $l=$3.4, 2.4, 2.8 ($\mu$m). (d--f) Phase-speed fitted by the three models. (g--i) $F(\theta)$ calculated using model (3). Other notations are the same as those of Fig. \ref{fig:simulation}.\label{fig:other examples}}
\end{figure}

Rotational movement of a single isolated cilium obtained from a ciliary mutant \textit{Dpcd} \cite{Kobayashi02} was recorded using high-speed imaging for 2 s (100 frames/s; HAS-500M; Ditect Co. Ltd.) \cite{Shinohara12}. The cilia tip portions were tracked manually to obtain 2-dim projection on the x--y plane (Fig. \ref{fig:simulation}(a), Figs. \ref{fig:other examples}(a--c)). The cilia setup and movement were characterized using the following parameters; the tilt angle of the rotation axis $\alpha$, the open angle $\beta$ of the rotation orbit along a conical face, the tilting direction $\eta$, cilium length $l$, and rotation phase $\theta$ (Fig. \ref{fig:ciliaparameter}). The phase $\theta$ is measured from the position closest to the cell surface. The projected trajectory to the observation plane ($x$-$y$) is elliptic as a result of tilting of the cilium. Therefore, $\theta$ was estimated as that after applying ellipse-correction \cite{Okada05} using multivariate nonlinear regression. Radii of the major and minor axes $a$ and $b$ were obtained from the fitted ellipse. The tilting direction $\eta$ was calculated from the minor axis of $b$ combining with the root position of cilia $\bf{R}$. The other parameters were calculated with the following relations derived from the construction shown in Fig. \ref{fig:ciliaparameter}(b, c):
	$\alpha=\cos^{-1}(b/a)$,
	$l=a\sqrt{1+\frac{s^2}{a^2-b^2}}$,
	$\beta=\sin^{-1}(a/l)$,
where $s$ denotes the distance between $\bf{R}$ and the center of the ellipse. Estimation of $\alpha$ includes some uncertainty (roughly assumed to be $\pm 10^\text{o}$ (degree) $=0.17$ (rad)) because of experimental setup problems: it is difficult to put the node samples perfectly flat against the observation plane.

To infer the driving force from the observed data, we used computer-aided fluid dynamics, where a cilium is modeled as a rigid cylinder with length of $l$ and diameter of $r$
rotating around a fixed orbit.
The driving force is balanced with the sum of the hydrodynamic drag force and the elastic restoring force. The magnitude of the elastic force must be almost zero or constant because curvature of the cilium is unchanged during the rotation as observed in the experiment \cite{Shinohara12}. Additionally the normal component of the drag force is much smaller than the tangential component. Consequently, the driving force can be approximated as the one balanced by the tangential component of the drag force. The equation of rotational movement in the Stokes flow regime is written as
\begin{equation}
	\label{eq:rotationEq}
	\dot{\theta}=\frac{F(\theta)}{J(\theta)},
\end{equation}
where $F(\theta)$ is the driving force required for the rotational motion, and $J(\theta)$ is the drag coefficient which is calculated numerically using boundary element method (BEM) \cite{Youngren75} as follows \footnote{Because $J(\theta)$ is the drag coefficient of the whole cilium, $F(\theta)$ is also the total driving force generated by the whole cilium}.

\begin{figure}
	\includegraphics[width=0.40\textwidth]{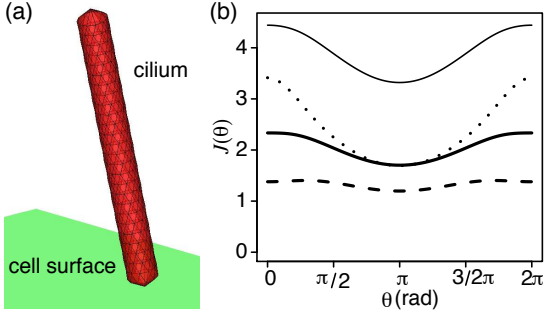}%
	\caption{ (color online) (a) Setup for boundary element method. Discretized cilium as a rigid cylinder (red). (b) Drag coefficient $J(\theta)$ normalized into dimensionless form by $\mu\omega l^2$, where $\mu$ is the viscosity, and where $\omega$ is the given rotational speed. The thick solid line shows the result obtained when the parameters was set as $l=1$, $\alpha=\pi/6$, $\beta=\pi/4$, $r=0.1$. Thin, dotted, and dashed lines respectively represent results when one parameter changed to $l=1.5$, $\alpha=2\pi/9$, and $\beta=2\pi/9$. \label{fig:dragfoef}}
\end{figure}

When a cilium moves in a fluid bounded by an infinite flat wall (approximation for a surface consisted of an array of cells), the flow field can be given as the following integral form \cite{pozrikidis92}:
\begin{equation}
	\bf u(\bf v)=-\frac{1}{8\pi\mu}\int_A\bf G(\bf v-\bf w){\bf q}({\bf w})dA,
	\label{eq:flowfieldEq}
\end{equation}
the vector $\bf v$ is a point where the velocity $\bf u$ is calculated, and $\bf w$ is a point where the traction force $\bf q$ is exerted,
$\mu$ represents the viscosity, and $A$ denotes the cilium surface.
The function $\bf G$ is the half-space Green function \cite{Blake74}, which rigorously satisfies the wall boundary condition \footnote{The component $ij$ of the Green function is defined with 
$G_{ij}=(\frac{\delta_{ij}}{r}+\frac{r_ir_j}{r^3})-(\frac{\delta_{ij}}{R}+\frac{R_iR_j}{R^3})+2h(\delta_{j\alpha}\delta_{\alpha k}-\delta_{jz}\delta_{zk})\frac{\partial}{\partial R_k}(\frac{hR_i}{R^3}-\frac{\delta_{iz}}{R}-\frac{R_iR_z}{R^3})$, where 
$\bf r=\bf v-\bf w$, $r=|\bf r|$, $\bf R=\bf v-\bf w'$, $R=|\bf R|$, $i, j, k = x, y, z$, $\alpha=x, y$, $\delta_{ij}$ is Kronecker delta, and $h$ is a distance from the wall $w_z=0$. The vector ${\bf w}' =(w_x, w_y, -h)$ is an image point of $\bf w$
}.
No background flow of the surrounding fluid is assumed. The NSB condition on the wall as well as the surface of cilia was assumed. The fluid has the same velocity to the rotating cilium on its surface. Using Eq. (\ref{eq:flowfieldEq}), one can calculate the traction force distribution on the cilium surface from the velocity distribution.
Equation (\ref{eq:flowfieldEq}) is solved numerically using BEM \cite{Ishikawa06}, where each cilium surface is discretized with 594 triangular elements (Fig. \ref{fig:dragfoef}(a)). The integration in Eq. (\ref{eq:flowfieldEq}) is performed on a triangle element using 28-point Gaussian polynomials. The singularity in the integration is solved analytically \cite{Youngren75}. The final linear algebraic equation is solved by LU decomposition, including pivoting. 
Because of the linearity of the Stokes flow, the drag force is proportional to the rotational velocity $\dot{\theta}$ \cite{Kim05}. Consequently, the strength of the drag force $F_d$ becomes equivalent to $J(\theta)$ when $\dot{\theta}=1$, where $F_d$ is defined as $F_d=-{\bf s_t}\cdot\int_{A}{\bf q}dA$. Therein, ${\bf s_t}$ is the unit tangential vector along the circular trajectory of the top of the cilium. Once Eq. (\ref{eq:flowfieldEq}) is solved for $\bf{q}$, $F_d$ can be calculated, so does $J(\theta)$.
In the numerical calculations, the parameters characterizing each cilium obtained from the experimental observations described above were used.

Figure \ref{fig:dragfoef}(b) presents examples of numerical results of $J(\theta)$. The drag force is maximized when the cilium passes closest to the cell surface ($\theta=0$). The mean value and the amplitude of the drag force increase when $\alpha$, $\beta$, or $l$ is enlarged. The mean values of $J(\theta)$ range 0.05--0.10 (pNs) as shown in Figs. \ref{fig:simulation}(c) and \ref{fig:other examples}(g--i). These values are slightly larger than those calculated from the relation on a drag coefficient$\frac{4/3\pi\mu l^2}{\ln (l/r)-0.447}$\footnote{the drag coefficients calculated using this form range 0.014--0.036 with the following parameters: $\mu=10^{-3}$ (Ns/m$^2$; 1.5-fold to 37$^\text{o}$C water viscosity), $l=2$--$3.5$ ($\mu$m), cilium diameter $r=0.4$ ($\mu$m).}, which is conventionally used for torque estimation of rotating molecular motors \cite{Hunt94}. In this relation, the NSB condition is considered, but a cilium is set parallel to the boundary. Therefore, tilting of the cilium is not considered. 


Now the numerical results of drag coefficients $J(\theta)$, and experimental observations of $\theta$ and $\dot{\theta}$ are combined by Eq. (\ref{eq:rotationEq}).
Using formulations $F(\theta)=F_0(1-F_r\cos(\theta-F_s))$ and $J(\theta-J_s)$ ( replacement of $J(\theta)$ with phase shift $J_s$ ),
three models are proposed as shown below. 
\begin{enumerate}[{(}1{)}]
	\item constant $F$ ($F_r=0$, $J_s=0$), 
	\item constant $F$ and shifted $J$ ($F_r=0$, $J_s \neq 0$) 
	\item variable $F$ depending on $\theta$ ($0 < F_r  < 1$, $J_s = 0$) 
\end{enumerate}
Here, phase shift $J_s $ was introduced to assess the possibility of estimation error of $\eta$, which might affect the definition for the position of $\theta=0$. Parameters $F_0$, $F_r$, and $F_s$ represent the mean driving force, force skewness ratio, and phase shift of force maximum from $\theta=\pi$ (the phase a cilium pass a position furthest from the cell surface), thereby $F_0 > 0$, $0 \le F_r \le 1$, and $-\pi \le F_s \le \pi$. Figure \ref{fig:simulation}(b, c) shows one application. The function $J(\theta)$ was calculated using BEM based on the experimental observation of $\alpha$, $\beta$, $l$ and $r=0.4$ ($\mu$m)
\footnote{The diameter of the cilium $r=0.4$ is estimated roughly from photographs of 9+0 cilium using electron microscopy \cite{Okada05, Chen11} for the range of 0.32--0.37 ($\mu$m), considering shrinkage resulting from vacuum conditions.}
\footnote{Data of the trajectory used for the estimation of $\alpha$, $\beta$, and $l$ are independent from the dataset of $\theta$ and $\dot{\theta}$ because the latter includes time.}.

To consult the degree of fitness statistically, model selection analysis was applied using the Akaike Information Criterion (AIC) \cite{Vries06}. The AIC value is a measure of the relative goodness of fit of a model, defined as $-2\ln L+2k$, where $L$ is the maximum likelihood function and $k$ is the number of parameters. Therefore, a smaller value of AIC reflects better fit of the model. Table \ref{table:aic} presents a summary of the result for the case of Fig. \ref{fig:simulation}. The estimation on $\alpha$ includes some uncertainty. Therefore, three variations (estimated $\alpha \pm 0.17$) were tested. Consequently, model (3) when $\alpha=0.37$ (No. 1) shows the best fit. 
Figure \ref{fig:other examples} shows other examples, and model (3) best explains the experimental data in any case \footnote{For the analysis presented above, the number of data $n=44$ for each dataset might not be sufficiently large to use AIC. Therefore, the datasets were also tested using the Bayesian Information Criterion 
and cross validation. The conclusions were almost identical to those obtained using AIC.}.
\begin{table}[h]
\caption{Model selection: AIC value for each model (1--3). The other parameters except $\alpha$ are the same as those in Fig. \ref{fig:simulation}
}
\begin{center}
\begin{tabular}{c|r||r|r|r}
	No. & $\alpha \text{(rad)}$ &  model (1) & model (2) & model (3) \\
\hline
	1 & 0.37 & 300.2 & 300.4 & \textbf{275.8}\\
\hline
	2 & 0.54 &  293.7 & 292.1 & 277.2\\
\hline
	3 & 0.72 &  287.7 & 280.7 & 278.2
	
\end{tabular}
\end{center}
\label{table:aic}
\end{table}%

\begin{figure}
	\includegraphics[width=0.48\textwidth]{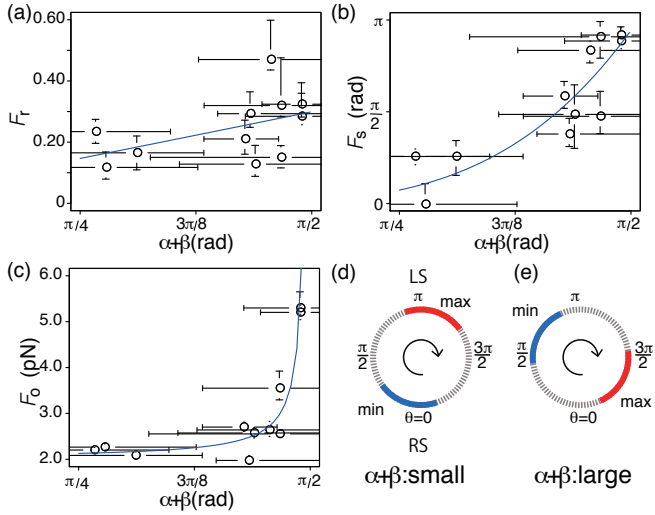}%
	\caption{(color online) (a--c) Relation between $\alpha + \beta$ and parameters of driving force $F(\theta)=F_0(1-F_r\cos(\theta -F_s ))$ estimated using model (3) from 11 trials obtained from four isolated cilia. 
	Vertical and horizontal error bars represent a 95\%- confidence interval associated, respectively, with the parameters of $F$ and with $\alpha+\beta$, which is estimated using the bootstrap method \cite{Efron86}, respectively, with the dataset of phase and phase speed, and ellipse fitting.
	Solid lines represent fitting by appropriate curves: $ax^b$ (a, b) and $a/(b-x)+c$ (c). The parameters were estimated using nonlinear regression: $a=0.19\pm0.05$, $b=1.03\pm0.73$ (a); $a=0.57\pm0.24$, $b=3.68\pm1.14$ (b); $a=0.096\pm0.048$ $b=1.57\pm0.01$, $c=2.01\pm0.23$ (c). (d,e) Schematic diagram for asymmetric driving force when $\alpha+\beta$ is small (d) and large (e).\label{fig:asymmetry}}
\end{figure}

Figure \ref{fig:asymmetry} summarizes estimated parameters $F_r$, $F_s$, and $F_0$ for $F(\theta)$ depending on $\alpha +\beta$ \footnote{Dependence of $F_r$, $F_s$, and $F_0$ on $\alpha$ or $\beta$ alone, and $l$ were also investigated. However, we found no correlation among them. }, of which the complementary angles correspond to the distance between a cilium and cell surface when the cilium passes closest to the cell surface. The force skewness ratio $F_r $ is 0.1--0.5s (Fig. \ref{fig:asymmetry}(a)), suggesting the existence of asymmetric stroke that cannot be explained only by the effect of tilting cilia with the NSB condition at the cell surface if a rigid straight cylinder for cilium can be assumed. Straightforwardly interpreting this result, the driving force is not constant and maximized at a phase $\theta=\pi+F_s$ (Fig. \ref{fig:asymmetry}(b)). In other words, it is maximized in LS and minimized in RS when $\alpha +\beta$ is small (up to $\sim 3\pi/8$) as shown in Fig. \ref{fig:asymmetry}(d), whereas it is maximized at the first half of RS when $\alpha +\beta$ close to $\pi/2$ 
 as shown in Fig. \ref{fig:asymmetry}(e).

The mean driving force $F_0$ was around 2 (pN) except when $\alpha$+$\beta$ is close to $\pi/2$. The order of magnitude is equivalent to the force generated using a single dynein head sliding along a microtubule in 9+2 cilia or flagella (estimated as 2--5pN in the literature) \cite{Hill10,Chen11}, which suggests that a single dynein participates in the generation of rotational movement at a moment. 

The value $F_0$ grows rapidly when $\alpha$+$\beta$ approaches $\pi/2$. The cilium skimmed over the cell surface in RS, then $F(\theta)$ maximized at around $3\pi/2\le\theta\le2\pi$ (Fig. \ref{fig:asymmetry}(e)). This can be understood from the observation that force generation of 9+2 cilia depends on a load \cite{Hill10}. This fact, however, does not explain how cilia accelerate in LS more than in the case expected by tilting cilia with NSB condition when $\alpha$+$\beta$ is much smaller than $\pi/2$ (Fig. \ref{fig:asymmetry}(d)). The following four ideas are conjectured as answers. First, rotational motion is regarded as generated by a dynein, protruding from one of nine doublets of microtubules attaching the next doublets sequentially in a circumferential direction in nine doubles \cite{Chen11}. The phase speed is greater in LS even if only tilting cilia with the NSB condition are considered, which can raise the probability of a dynein attaching to the next doublet. This attachment can accelerate the rotation speed in LS and vice versa in RS. Second, tensions acting on certain cytoskeleton connected to a basal body can be asymmetric \footnote{the basal body is an apparatus in which the axoneme (9+0 or 9+2 microtubules) continues down into the cell} because of shifting of the basal-body position in each cell (Fig. \ref{fig:nodalcilia}(b)) \cite{Hashimoto10}. Third, axonemes themselves can be asymmetric. Such a structure is found in 9+2 cilia \cite{Ueno12}. Finally, the arcuate shape of cilia can also affect the asymmetric stroke \cite{Ikegami10}.
 
In conclusion, our results suggest the existence of asymmetric strokes that cannot be explained adequately by the effect of cilia tilting with NSB condition at the cell surface. This would result from the asymmetric force generation in LR strokes if a rigid straight cylinder for cilium can be assumed. In an early stage of development, the number of cilia, length, rotational speed, and tilting angles are small \cite{Shinohara12, Hashimoto10}. However, it was revealed recently that sufficient flow speed for LR determination can be achieved by only two cilia \cite{Shinohara12}. Under such an unreliable situation by a few cilia, the asymmetric stroke can support the establishment of LR determination more effectively.
 
\begin{acknowledgments}
We thank Dr. N. Motoyama, NCGG, Japan for providing \textit{dpcd} and thank MEXT Japan for financial support.
\end{acknowledgments}


%

\end{document}